\documentclass[twocolumn]{webofc}
\usepackage[varg]{txfonts}   % Web of Conferences font

\usepackage{pbox,calc}
\usepackage{subfigure}
\usepackage{mathtools}
\usepackage{amsmath}
\usepackage{amssymb}
\usepackage{amsmath}
\usepackage{stfloats}

\usepackage[compact]{titlesec}
\begin{document}

\title{In search of the best nuclear data file for proton induced reactions: varying both models and their parameters}

\author{\firstname{E.} \lastname{Alhassan}\inst{1,2}\fnsep\thanks{\email{erwin.alhassan@psi.ch}} \and
        \firstname{D.} \lastname{Rochman}\inst{1}\fnsep\thanks{\email{dimitri-alexandre.rochman@psi.ch}} \and
        \firstname{A.} \lastname{Vasiliev}\inst{1}\and
         \firstname{R.M.} \lastname{Bergmann}\inst{1}\and
          \firstname{M.} \lastname{Wohlmuther}\inst{2}\and
          \firstname{A.J.} \lastname{Koning}\inst{3,4}\and     
          \firstname{H.} \lastname{Ferroukhi}\inst{1}                   
        % etc.
}

\institute{Laboratory for Reactor Physics and Thermal-Hydraulics, Paul Scherrer Institute, 5232 Villigen, Switzerland
\and
           Division Large Research Facilities (GFA), Paul Scherrer Institute, Villigen, Switzerland
\and
           Nuclear Data Section, International Atomic Energy Commission (IAEA), Vienna, Austria
\and
           Division of Applied Nuclear Physics, Department of Physics and Astronomy, Uppsala University, Uppsala, Sweden
          }

\abstract{%
A lot of research work has been carried out in fine tuning model parameters to reproduce experimental data for neutron induced reactions. This however is not the case for proton induced reactions where large deviations still exist between model calculations and experiments for some cross sections. In this work, we present a method for searching both the model and model parameter space in order to identify the 'best' nuclear reaction models with their parameter sets that reproduces carefully selected experimental data. Three sets of experimental data from EXFOR are used in this work: (1) cross sections of the target nucleus (2) cross sections of the residual nuclei and (3) angular distributions. Selected models and their parameters were varied simultaneously to produce a large set of random nuclear data files. The goodness of fit between our adjustments and experimental data was achieved by computing a global reduced chi square which took into consideration the above listed experimental data. The method has been applied for the adjustment of proton induced reactions on $^{59}$Co between 1 to 100 MeV. The adjusted files obtained are compared with available experimental data and evaluations from other nuclear data libraries. 
}
\maketitle
\section{Introduction}
\label{intro}
High quality proton nuclear data are important for a wide range of applications, e.g., in proton therapy, medical radioisotope production, accelerator physics as well as in astrophysics, for a better understanding of stellar nucleosynthesis, among others. Similar to neutrons, the evaluation of proton induced reactions normally involves a combination of nuclear reaction modelling and carefully selected experimental data. Despite the progress made in nuclear reaction theory over the past decade, comparison of model calculations with experimental data usually reveals discrepancies between the two. A common solution is to adjust or fine tune parameters to nuclear reaction models in order to fit differential experimental data obtained from the EXFOR database~\cite{EXFOR-2007}. 

A single nuclear reaction calculation involves several models with several parameters, linked together in a nuclear reaction code such as TALYS~\cite{TALYS-2007} or EMPIRE~\cite{Herman-2007empire}. In the TALYS code for example, there are six level density models, three optical models, four pre-equilibrium models and eight gamma-strength models, among others implemented. A combination of these models usually gives different TALYS outputs. One often overlooked but important step in the evaluation process is the identification of model combinations that can reproduce experimental data. This is in part due to the fact that, for several decades, much effort driven largely by the reactor community has been put into improving the neutron-sub library through the identification and fine tuning of model parameters for a large number of isotopes in the case of the TENDL library~\citep{Tendl-2017} for example. The identified models have been used over the years for evaluations without necessarily going back to the model selection step. In Ref.~\cite{Koning-2019tendl} however, it was demonstrated that, the simultaneous variation of models and their parameters induces prior correlations and therefore could have significant impact on nuclear data adjustments. In Ref.~\cite{Koning-2019tendl}, model selection and the adjustment of models (and their parameters) in order to fit differential experimental data was not emphasized. Until recently, much effort was not devoted to the evaluation of proton induced reactions which is evident by the number of evaluations available in the proton sub-library in the major nuclear data libraries compared with the neutron sub-library: 49 isotopes in the ENDF/B-VIII.0 library and 106 in the JENDL/HE-2007 (JENDL High Energy file) library compared with 557 isotopes in the neutron sub-library for ENDF/B-VIII.0, 406 for JENDL-4.0 and 562 for the JEFF-3.3 library. In the case of the proton induced reactions, the TENDL-2017 and JEFF-3.3 libraries both contain evaluated nuclear data for 2804 isotopes, however, the evaluations were all carried out with default TALYS models and parameters~\cite{Koning-2019tendl}. Furthermore, both the model and parameter space in the case of proton induced reactions have been left largely unexplored, necessitating for the simultaneous variation of both models and their parameters as proposed in this work.

\section{Method}
\label{sec-1}
A total of 200 random model combinations were generated by varying a selected number of nuclear reaction models implemented within the TALYS code. These model combinations were run with the TALYS code (version 1.9) to produce a large set of random physical observables referred here as the \emph{parent generation}. A total of 682 random nuclear data were produced for the parent generation. The parent generation as used in this work refers to the initial random nuclear data (ND) files generated from the variation of models. 

The random nuclear data files in the ENDF format were processed into XY tables for comparison with selected experimental data from the EXFOR database using a reduced $\chi^2$. Based on the $\chi^2$, the model combination with the minimum $\chi^2$ was chosen as the 'best' model set. The selected model combination (also referred to as the parent file) was used as the nominal file for re-sampling of model parameters to produce the next generation of TALYS outputs (referred to as the 1st generation). The output of the 1st generation were again compared with experimental data from the EXFOR database and new 'best' file was selected. 

\subsection{Experimental data used}
Three experimental categories were used: (1) cross sections of the target nucleus (2) cross sections of the residual nuclei (also called the residual production cross sections) and (3) angular distributions. In the case of cross sections of the target nucleus (also referred to as the reaction cross sections in this work), the following eight channels were considered in the adjustments: (p,non-el), (p,n), (p,3n), (p,4n), (p,2np)g, (p,2np)m, (p,$\gamma$) and (p,xn) and for the residual production cross sections: $^{59}$Co(p,x)$^{46}$Sc, $^{59}$Co(p,x)$^{48}$V, $^{59}$Co(p,x)$^{52}$Mn, $^{59}$Co(p,x)$^{55}$Fe, $^{59}$Co(p,x)$^{55}$Co, $^{59}$Co(p,x)$^{56}$Co, $^{59}$Co(p,x)$^{57}$Co, $^{59}$Co(p,x)$^{58}$Co, $^{59}$Co(p,x)$^{57}$Ni. In the case of angular distributions, only the elastic angular distributions were considered. 

A total of 169, 141 and 185 experimental data points were used for the reaction cross sections, the residual production cross sections and the elastic angular distributions respectively. Similar to Ref.~\cite{Alhassan-2018Nmultilevel}, experiments that were observed to be inconsistent with other experimental sets and deviate from the trend of our model calculations as well as other evaluations (when available), were not considered. Also, for the cases where the only experimental data available for a particular energy range has no uncertainties reported, we assume a 10\% uncertainty for that experimental set.

\subsection{Optimization of models and their parameters to experimental data} 
In this work, the reduced $\chi^2$ was used as the goodness of fit estimator. Since three experimental categories were used in the adjustment, we take all these experimental data into account by computing a global $\chi^2$ given as follows: 
\begin{equation}
\chi_{G,k}^2 = \chi_{k}^2(xs) + \chi_{k}^2(rp) + \chi_{k}^2(DA) 
\label{global_chi2}
\end{equation}

where $\chi_{G,k}^2 $ is the global chi square for the random nuclear data $k$, $\chi_{k}^2(xs)$ and $\chi_{k}^2(rp)$ are the chi squares computed using the reaction cross sections and the residual production cross sections respectively, and $\chi_{k}^2(DA)$ is the chi square computed for the elastic angular distributions. For Eq.~\ref{global_chi2} to hold, it was assume that the different experimental categories as presented were uncorrelated and were of equal importance in the adjustment. Further, similar to Refs.\cite{koning-2015bayesianfull,Alhassan-2018Nmultilevel}, the experimental data points were assumed to be uncorrelated. The reason being that, experimental correlations especially for proton induced reactions were not readily available. Our reduced $ \chi^2_{c(k)}$ for the channel $c$ and nuclear data (ND) file $k$, can be given as:

\begin{equation}
    \chi^2_{c(k)} = \frac{1}{N_p} \sum_{i=1}^{N_p} \bigg(  \frac{\sigma^i_{T(k)} - \sigma^i_E}{\Delta \sigma^i_E} \bigg)^2
    \label{gen_chi2}
\end{equation}

where $\sigma^i_{T(k)}$ is a vector of TALYS calculated observables found in the $k^{th}$ random ND file for the channel $c$ and $\sigma^i_E$ is a vector of experimental observables as a function of incident neutron energy ($i$) for channel $c$,  $\Delta \sigma^i_E$ is the experimental uncertainty at an incident energy $i$ of channel $c$, and $N_p$ is the total number of experimental points per reaction channel considered. In cases where no matches in energy ($i$) were observed between the TALYS output obtained and the experimental data for the $c^{th}$ channel, we carry out a linear interpolation in order to fill in the missing TALYS values. In the case of angular distributions, only the missing values in angle were filled through linear interpolation. In order to obtain perfect matches in energy for the elastic angular distributions, the energies at which angular distributions where measured where given to the TALYS code as input. From Eq.~\ref{gen_chi2}, the reduced chi square for the reaction cross section ($\chi_{k}^2(xs)$) for example, can be given as:
\begin{equation}
    \chi_{k}^2(xs) = \frac{1}{N_c} \sum_{c=1}^{N_c} \chi^2_{c(k)}
    \label{chi2_chan}
\end{equation}
where $N_c$ is the number of considered channels. In Ref.~\cite{Alhassan-2018Nmultilevel}, a weighted $\chi^2$ where channel weights proportional to the average channel cross section, was presented. The idea was to assign channels with large average cross sections higher weights and those with lower relatively smaller average cross sections, lower weights. However, since the goal of this work is to produce a TENDL based evaluation for a general purpose library, all channels were assumed to carry equal weights. The file with the minimum global $\chi^2$ (with its set of models) was selected as our best file and used as the nominal file around which model parameters were varied. Because of computational resource constraints, the final 'best' file produced was based on the results of the 1st generation. Also, for the selection of models, the Bayesian approach for model selection could have been used. This approach is presented in a dedicated paper~\cite{Alhassan-2019Modelselect}. 

\section{Results and Discussion}
\begin{figure*}[b!] 
  \center
  \includegraphics[trim = 20mm 70mm 20mm 70mm, clip, width=0.85\textwidth]{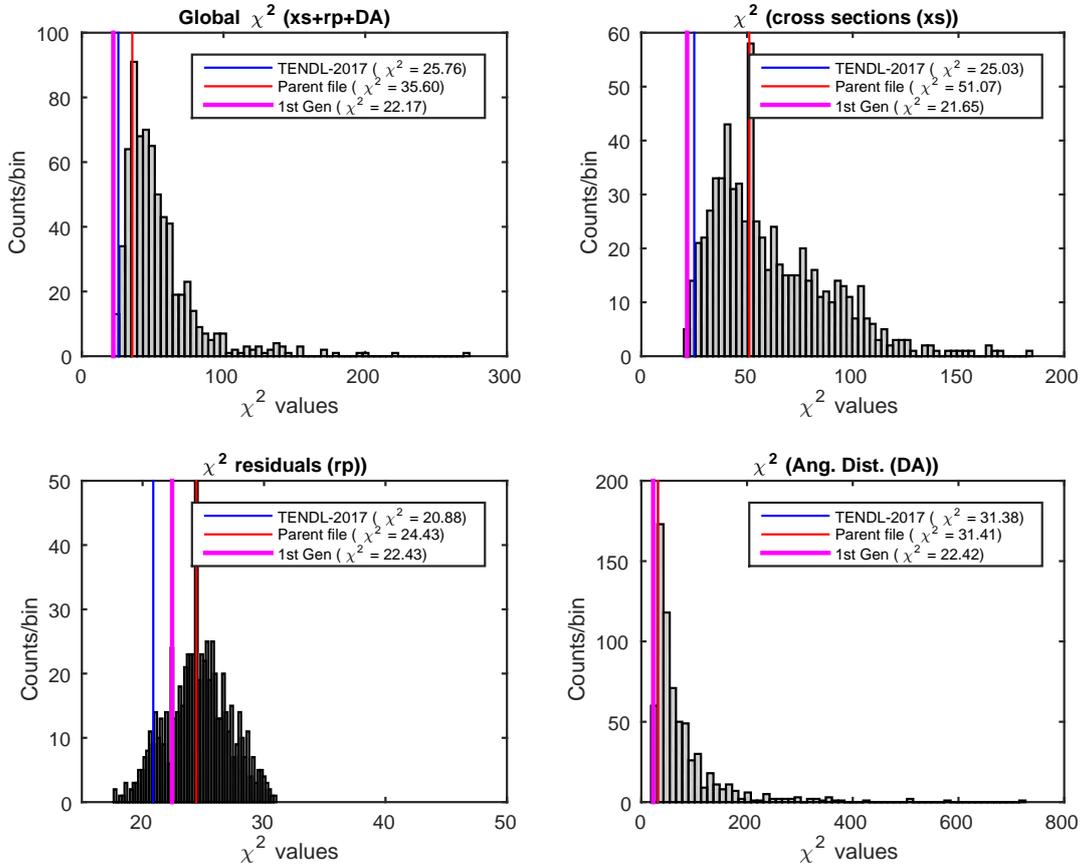} 
  \caption{$\chi^2$ distributions for the 1st generation for the three experimental data categories as well as the global $\chi^2$ are presented. xs denotes reaction cross sections, rp – residual production cross sections and DA – angular distributions. A total of 682 random samples were used for each plot.}
  \label{chi_sq_distr}
  \end{figure*}

In Fig.~\ref{chi_sq_distr}, the global $\chi^2$ distribution as well as the $\chi^2$ distributions for the reaction cross sections (xs), the residual production cross sections (rp) and the angular distributions (DA) for the 1st generation are presented and compared with $\chi^2$ values computed for the TENDL-2017 library and the 'best' file from parent generation (referred to as the 'parent file'), using the same experimental data.  
From Fig.~\ref{chi_sq_distr}, it can be seen that, the adjustment from the 1st Gen out performed the TENDL-2017 evaluation for the reaction cross sections and the angular distributions but performed quite poorly with respect with to the residual production cross sections. Also, it can be observed that, the results from the 1st generation is an improvement over the parent file as expected: $\chi^2$ values of 22.17, 21.65, 22.43, 22.42 for the global, reaction cross sections, residual production cross sections and angular distributions respectively, were obtained for the 1st Gen compared with 35.60, 51.07, 24.43, and 22.42 for the parent file. To improve on the 1st generation, the new 'best' file  obtained could have been used as the nominal for re-sampling of model parameters in an iterative fashion. This however, can be computationally expensive and therefore not carried out in this work. 

\begin{figure*}%[h!] %tb]
  \centering
  \includegraphics[trim = 15mm 20mm 10mm 18mm, clip, width=0.40\textwidth]{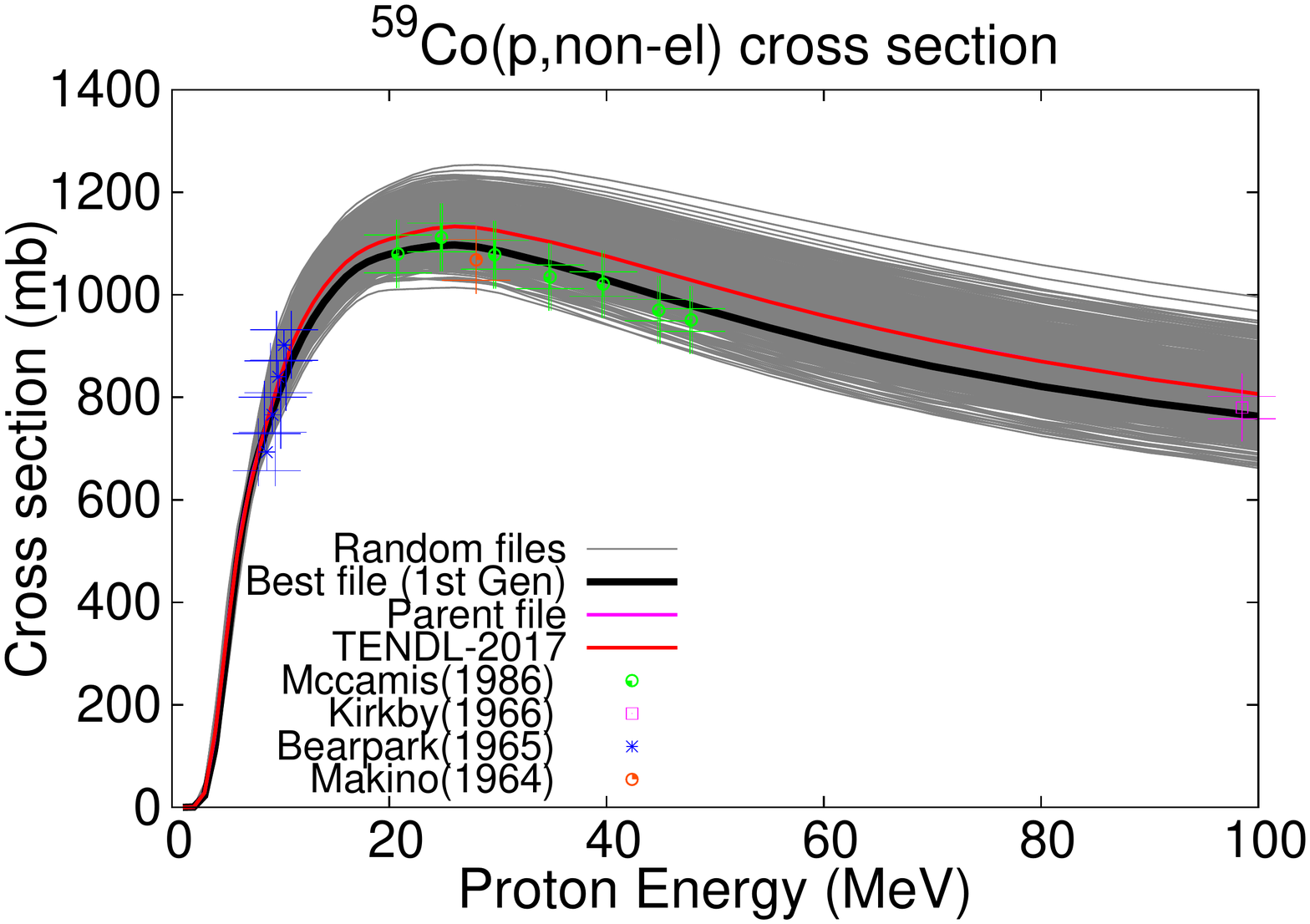}
  \includegraphics[trim = 15mm 20mm 10mm 18mm, clip, width=0.40\textwidth]{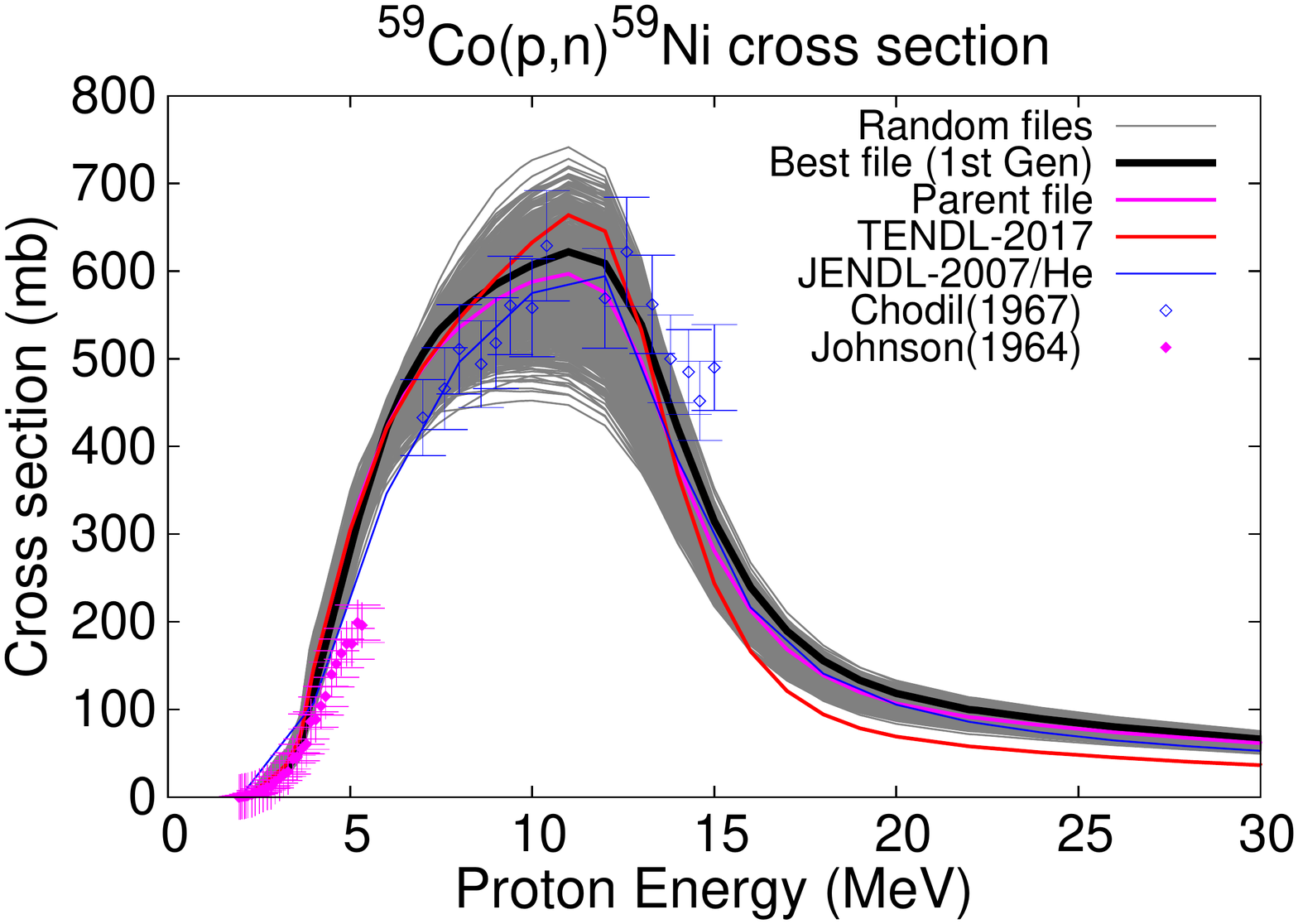}
  \caption{Comparison of file performance between the evaluations from this work and the TENDL-2017 library for the (p,non-el), (p,n) cross sections of $^{59}$Co. Comparisons are made with the JENDL/He-2007 library in cases where evaluations are available.  Only the experimental data sets used in the adjustment have been presented.}
  \label{file_performance_xs}
  \end{figure*}

In Fig.~\ref{file_performance_xs}, a comparison of file performance between our evaluations and the TENDL-2017 library are presented for the (p,non-el) and (p,n) cross sections of $^{59}$Co. In cases where evaluations are available, comparisons are made also with the JENDL-2007/He library. From the figure, it can be observed that, the evaluation from the 1st generation performed better than the TEND-2017 library for the (p,non-el) and (p,n) cross sections. The TENDL-2017 evaluation over estimates the (p,non-el) cross section from about 20 to 100 MeV while this evaluation is within the experimental uncertainties over the entire incident energies. 

\begin{figure*}%[h!] %tb]
  \centering
  \includegraphics[trim = 15mm 18mm 10mm 20mm, clip, width=0.40\textwidth]{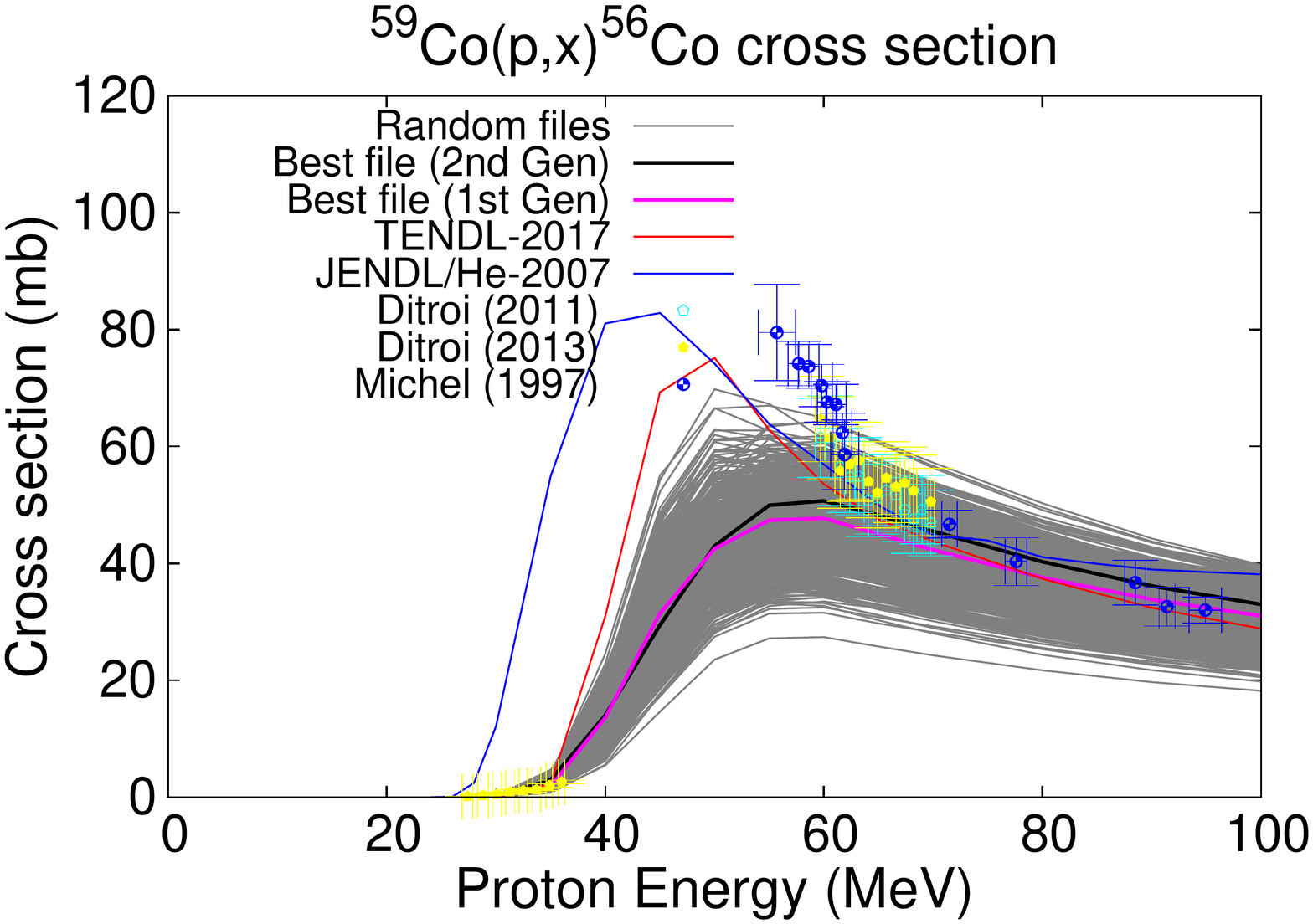}
  \includegraphics[trim = 15mm 18mm 10mm 20mm, clip, width=0.40\textwidth]{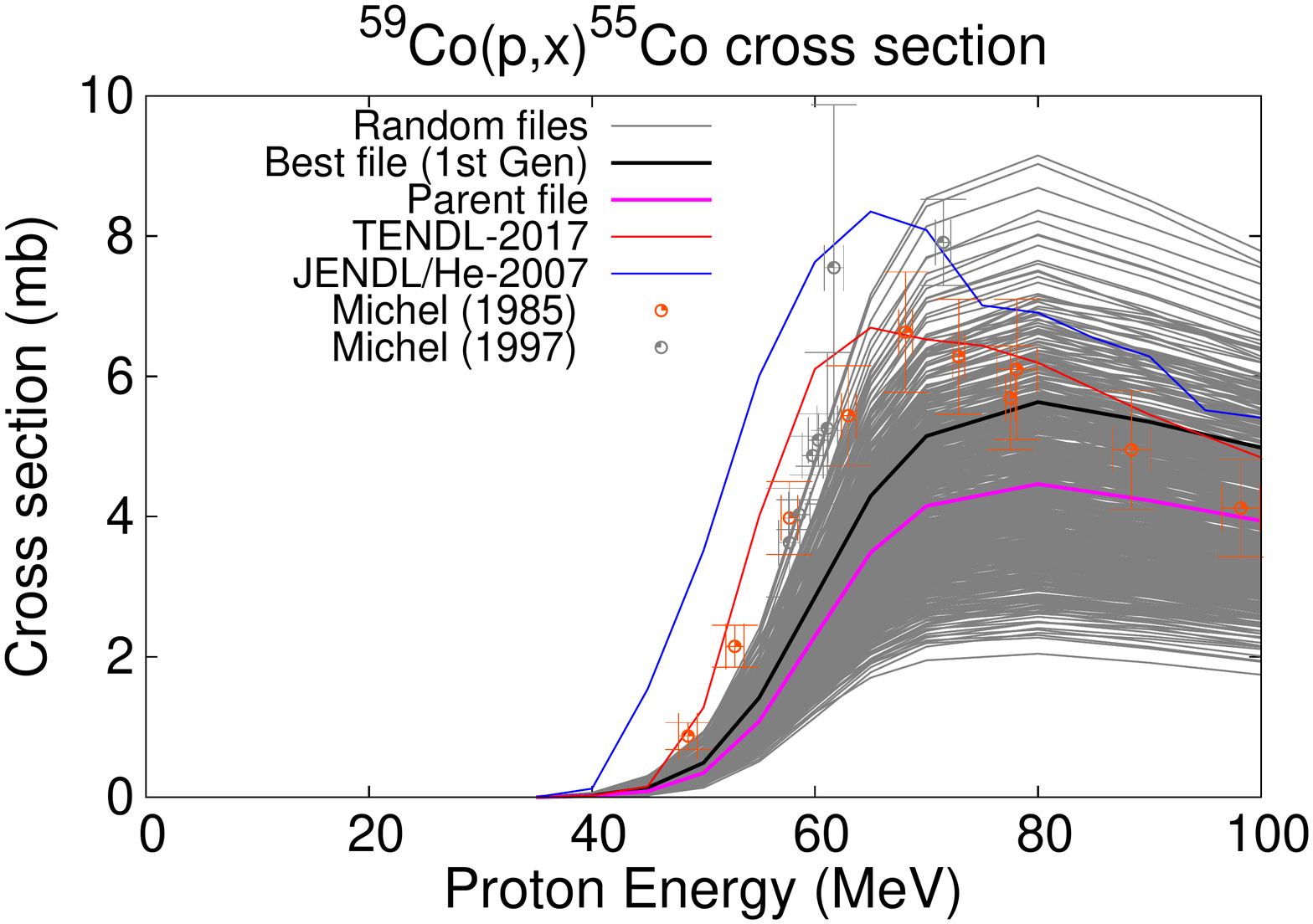}
  \caption{Comparison of file performance between our evaluation and the TENDL-2017 evaluation as well as the JENDL/He-2007 for the $^{59}$Co(p,x)$^{56}$Co and $^{59}$Co(p,x)$^{55}$Co residual production cross sections.}
  \label{file_performance_rp}
  \end{figure*}

Fig.~\ref{file_performance_rp} presents the comparison of file performance between our evaluation, the TENDL-2017 and JENDL/He-2007 evaluations for the $^{59}$Co(p,x)$^{56}$Co and $^{59}$Co(p,x)$^{55}$Co residual production cross sections. In the case of the $^{59}$Co(p,x)$^{56}$Co for example, our evaluation (i.e. the 1st Gen), under predicts the data at incident energies below 60 MeV. Our evaluation however describes the experimental data reasonably well between 60 to 100 MeV. Similarly in the case of the $^{59}$Co(p,x)$^{55}$Co, our evaluation is unable to fit satisfactorily to experimental data. This explains the relatively large $\chi^2$ value of 22.43 obtained for this evaluation (1st Gen) compared with 20.88 obtained for TENDL-2017 with respect to the residual production cross sections. In order to improve the residual cross sections, the 'best' file from the 1st Gen can be utilized as the new nominal file for parameter variation in an  iterative fashion. This is however planned for future work. 

\section{Conclusion}
%\linespread{1.0}
A method was presented for searching the model and parameter space through the simultaneous variation of many TALYS models (and their parameters). By computing a reduced global $\chi^2$ which takes into consideration experimental information from reaction and residual production cross sections as well as the elastic angular distributions, we were able to identify a file that performs favourably globally when compared with the TENDL-2017 evaluation. The method has been applied for the adjustment of proton induced reactions on $^{59}$Co from 1 to 100 MeV. It was observed that, by exploring a larger model space, model combinations that reproduce differential experimental data can be identified for the model parameter variation step. The study also shows that there is a potential for improvement of  evaluations (within the limit of the models), through an iterative process. 
%
% \medskip
%
% BibTeX or Biber users please use (the style is already called in the class, ensure that the "woc.bst" style is in your local directory)
% \bibliography{name or your bibliography database}

\begin{thebibliography}{}
\bibitem{EXFOR-2007}
H. Henriksson, O. Schwerer, D. Rochman, M. Mikhaylyukova, and N. Otuka.
International Nuclear Data Conference for Science and Technology, Nice, France, April, 22-27 (2007).
%Journal Author, Journal \textbf{Volume}, page numbers (year)

\bibitem{TALYS-2007}
A.J. Koning, S. Hilaire, and M.C. Duijvestijn.
Nuclear Data Conference for Science and Technology, Nice, France, April, 22-27 (2007).

\bibitem{Herman-2007empire}
M. Herman, R. Capote, B.V. Carlson, P. Oblo{\v{z}}insk{\`y}, M. Sin, A. Trkov, H. Wienke, and V. Zerkin, Nuclear Data Sheets \textbf{108}, 2655-2715 (2007).

\bibitem{Tendl-2017}
D. Rochman, A.J. Koning, J.C. Sublet, M. Fleming, E. Bauge, S. Hilaire, P. Romain, B. Morillon, H. Duarte, S. Goriely, S.C. van der Marck, H. Sj{\"o}strand, S. Pomp, N. Dzysiuk, O. Cabellos, H. Ferroukhi, and A. Vasiliev, EPJ Web of Conferences \textbf{146},  02006 (2017).

\bibitem{Koning-2019tendl}
A.J. Koning, D. Rochman, J. C. Sublet, N. Dzysiuk, M. Fleming, and S.C van der Marck, Nuclear Data Sheets \textbf{155}, 1-55 (2019).

\bibitem{Ditroi-2011Co}
F. Ditr{\'o}i, S. Tak{\'a}cs, F. T{\'a}rk{\'a}nyi, R.W. Smith, and M. Baba,  J. Korean Phys Society \textbf{59}, 1697-1700 (2011).

\bibitem{Alhassan-2018Nmultilevel}
E. Alhassan, D. Rochman, H. Sj\"ostrand, A. Vasiliev,  A.J. Koning, and H. Ferroukhi, Bayesian updating for data adjustments and multi-level uncertainty propagation within Total Monte Carlo, Under review in Annals of Nuclear Energy (2019).

\bibitem{koning-2015bayesianfull}
A.J. Koning,  The European Physical Journal A \textbf{51}, 1-16 (2015).

\bibitem{Alhassan-2019Modelselect} 
E. Alhassan, D. Rochman, A. Vasiliev,  M. Wohlmuther, A.J. Koning, and H. Ferroukhi, Model selection for nuclear data adjustments and evaluation, In Manuscript (2019).
\end{thebibliography}

\end{document}